# Dark Matter Search Perspectives with GAMMA-400

A.A. MOISEEV[1], A.M. GALPER[2,3], O. ADRIANI[4], R.L. APTEKAR[4][5], I.V. ARKHANGELSKAJA[3], A.I. ARKHANGELSKIY[3], G.A. AVANESOV[6], L. BERGSTROM[7], M. BOEZIO[8], V. BONVICINI[8], K.A. BOYARCHUK[9], V.A. DOGIEL[2], YU.V. GUSAKOV[2], M.I. FRADKIN[2], CH. FUGLESANG[10], B.I. HNATYK[11], V.A. KACHANOV[12], V.A. KAPLIN[3], M.D. KHEYMITS[3], V. KOREPANOV[13], J. LARSSON[10], A.A. LEONOV[3], F. LONGO[8], P. MAESTRO[14], P. MARROCCHESI[14], E.P. MAZETS[5], V.V. MIKHAILOV[3], E. MOCCHIUTTI[8], N. MORI[4], I. MOSKALENKO[15], P.YU. NAUMOV[3], P. PAPINI[4], M. PEARCE[10], P. PICOZZA[17], M.F. RUNTSO[3], F. RYDE[10], R. SPARVOLI[17], P. SPILLANTINI[4], S.I. SUCHKOV[2], M. TAVANI[17], N.P. TOPCHIEV[2], A. VACCHI[8], E. VANNUCCINI[4], YU.T. YURKIN[3], N. ZAMPA[8], V.N. ZARIKASHVILI[18], V.G. ZVEREV[3] FOR THE GAMMA-400 COLLABORATION

[1] *CRESST/GSFC and University of Maryland, College Park, Maryland 20742, USA*
[2] *Lebedev Physical Institute, Russian Academy of Sciences, Moscow, Russia*
[3] *National Research Nuclear University MEPhI, Moscow, Russia*
[4] *Istituto Nazionale di Fisica Nucleare, Sezione di Firenze and Physics Department of University of Florence, Firenze, Italy*
[5] *Ioffe Physical Technical Institute, Russian Academy of Sciences, St. Petersburg, Russia*
[6] *Space Research Institute, Russian Academy of Sciences, Moscow, Russia*
[7] *Stockholm University, Department of Physics; and the Oskar Klein Centre, AlbaNova University Center, Stockholm, Sweden*
[8] *Istituto Nazionale di Fisica Nucleare, Sezione di Trieste, Trieste, Italy*
[9] *Research Institute for Electromechanics, Istra, Moscow region, Russia*
[10] *KTH Royal Institute of Technology, Department of Physics; and the Oskar Klein Centre, AlbaNova University Center, Stockholm, Sweden*
[11] *Taras Shevchenko National University of Kyiv, Ukraine*
[12] *Institute for High Energy Physics, Protvino, Moscow region, Russia*
[13] *Lviv Center of Institute of Space Research, Lviv, Ukraine*
[14] *Istituto Nazionale di Fisica Nucleare, Sezione di Pisa and Physics Department of University os Siena, Siena, Italy 14*
[15] *15 Hansen Experimental Physics Laboratory and Kavli Institute for Particle Astrophysics and Cosmology, Stanford University, Stanford, USA*
[16] *16 Istituto Nazionale di Fisica Nucleare, Sezione di Roma 2 and Physics Department of University of Rome Tor Vergata, Rome, Italy*
[17] *17 Istituto Nazionale di Astrofisica IASF and Physics Department of University of Rome Tor Vergata, Rome, Italy*
[18] *18 Pushkov Institute of Terrestrial Magnetism, Ionosphere, and Radiowave Propagation, Troitsk, Moscow region, Russia*

*alexander.a.moiseev@nasa.gov*

**Abstract:** GAMMA-400 is a future high-energy $\gamma$-ray telescope, designed to measure the fluxes of $\gamma$-rays and cosmic-ray electrons + positrons, which can be produced by annihilation or decay of dark matter particles, and to survey the celestial sphere in order to study point and extended sources of $\gamma$-rays, measure energy spectra of Galactic and extragalactic diffuse $\gamma$-ray emission, $\gamma$-ray bursts, and $\gamma$-ray emission from the Sun. GAMMA-400 covers the energy range from 100 MeV to $\sim$3000 GeV. Its angular resolution is $\sim 0.01°(E_\gamma > 100$ GeV), and the energy resolution $\sim 1\ \%\ (E_\gamma > 10$ GeV). GAMMA-400 is planned to be launched on the Russian space platform Navigator in 2019. The GAMMA-400 perspectives in the search for dark matter in various scenarios are presented in this paper

**Keywords:** $\gamma$-ray telescope, $\gamma$-ray observations, particle dark matter

## 1 Introduction

The GAMMA-400 $\gamma$-ray telescope is designed to measure the cosmic $\gamma$-ray, electron + positron and nuclei fluxes in a wide energy range. The energy and arrival direction measurement will be of unprecendented accuracy for such experiments, and charged particle background rejection in $\gamma$-ray measurement will be very high due to the presence of a deep calorimeter. The objectives of the mission and its detailed description are given in [1] and [2], and the most current status of the project with recent modification is given in [3]. The main target for this mission is to conduct a sensitive search for signatures of particle dark matter (DM) in high energy diffuse $\gamma$-radiation. This task was set for this project by Nobel Laureate Academician V. L. Ginzburg in the end of 1980's [4, 5] and his list of very important issues in modern cosmology at the beginning of XXI Century noted the issue of dark matter and its detection [6].

Currently successfully operating for more than 5 years, the high-energy $\gamma$-ray space telescope Fermi LAT [7] is conducting a deep survey of many possible scenarios in high-energy $\gamma$-ray radiation, which may be associated with annihilation or decay of dark matter particles, e.g. WIMPS. GAMMA-400 is very suitable for the search for WIMPs and due to its enhanced performance at energies above 10 GeV (point spread function and energy resolution) will follow and deepen the Fermi LAT findings.

## 2 Dark Matter Search Scenarios

Indirect searches for dark matter in cosmic $\gamma$-rays are often based on the search for disagreement or difference be-tween observed $\gamma$-ray flux, spectrum and spatial distribution and that predicted by classical models. GAMMA-400 will conduct the search in the phase space of $\gamma$-rays with energy above $\sim 1$ GeV and electron + positron spectrum above $\sim 10$ GeV. The general directions of the search are as follows: a) investigating the discreapancy between luminous and gravitational matter observed in some objects such as dwarf spheroidal galaxies (dSph) and galaxy clusters; b) accurate measurement of the $\gamma$-ray spectrum from DM halo and comparing it to the "conventional" diffuse background spectrum; c) search for $\gamma$-ray sources (satellites), likely extended, that do not have association with sources in other wavelength radiation; d) search for $\gamma$-ray lines in the continuum spectrum; e) analysis of the composition of extragalactic $\gamma$-background radiation if there is an unaccounted fraction; f) study of the fine structure of high-energy electron spectrum (Fig.1). One of the decisive steps in identifying a DM signal in spectral study is to distinguish it from that originated in astrophysical objects, unless it is a monochromacic line (see next section). It is likely that WIMP annihilation or decay produces a relatively harder continuum spectrum than that from astrophysical objects, with a bump or edge near the WIMP mass, and this difference can be a criterion for detection (see e.g. review [8]). The results of our study will be interpreted within a variety of DM frameworks, including the most popular minimal supersymmetric standard model (MSSM) and universal extra dimensions (EUD).

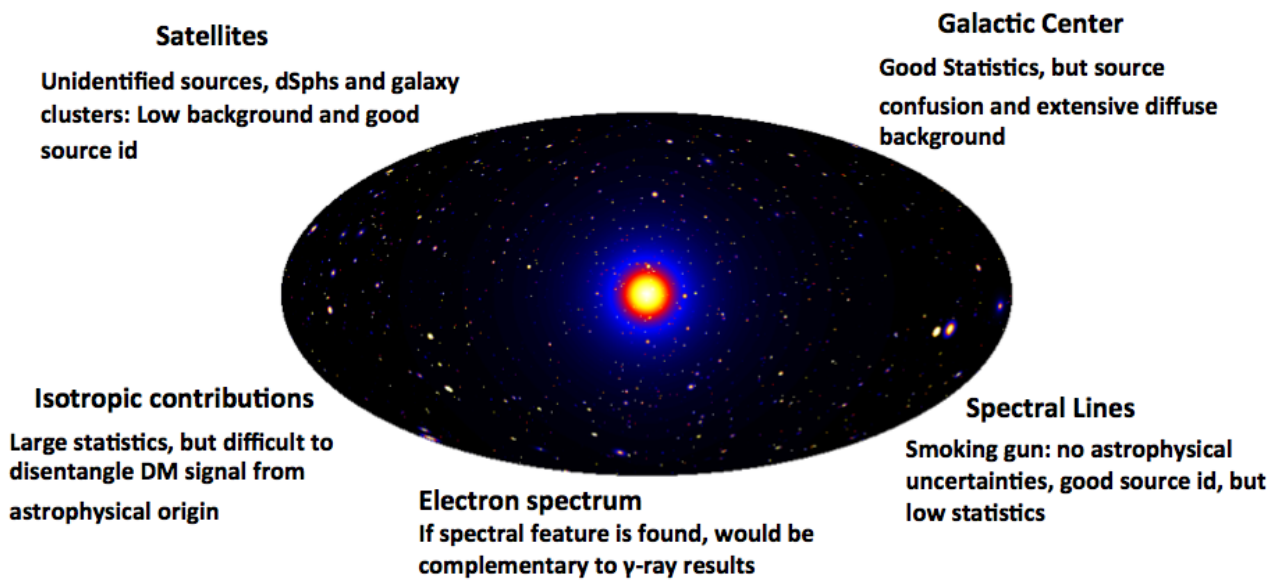

**Figure 1**: Possible scenarios in indirect dark matter search in $\gamma$-rays

Fermi LAT made significant progress in indirect DM search, focused on a search for $\gamma$-ray lines, study of radiation from dSph and galaxy clusters, search for satellites among unidentified $\gamma$-sources, and accurate measurement of diffuse spectrum. GAMMA-400 will follow these directions in its search for DM. The sensitivity to a DM signal depends critically on accurate knowledge of the backgrounds: diffuse Galactic and isotropic $\gamma$-rays, $\gamma$-rays from astrophysical sources, and charged particles detected as $\gamma$-rays. The assessment of these backgrounds will be the special task for the GAMMA-400 mission.

### 2.1 Search for $\gamma$-ray lines

A "smoking gun" in the DM search would be the detection of monochromatic $\gamma$-ray line on top of the continuum background spectrum. With high confidence it would be interpreted as a line emission in WIMP annihilation into two photons, which for slowly moving dark matter particles would give rise to a striking, almost monoenergetic as $\delta$-function photon signal. Its spatial shape is expected to be extended up to $0.1^0 - 0.3^0$, however point sources are not excluded either.

Probably the most exciting recent result in the indirect search for dark matter is the Fermi LAT "135 GeV" line. First reported in [9] (Fig.2), it has been widely discussed by scientists of different fields. Most accurate current status is given in the Fermi LAT paper [10], with the conclusion that the global significance of the feature is 1.6 $\sigma$, which is insufficient to claim the discovery; however more work is needed to understand what it is. The line itself has higher local significance ($\sim 3.5\sigma$), but after careful consideration of possible effect of systematic errors the net global significance reduces to 1.6 $\sigma$. Fermi LAT will continue extensive efforts focused on this extremely important result. GAMMA-400, with its high energy and angular resolution at high energy, is the only currently planned experiment which has very good perspectives to resolve the situation around this potential discovery. If the line originates in point source, GAMMA-400 will be >5 times more sensitive to detect it. However, a more likely scenario is that the line is produced in extended sources, likely at Galactic center. In this case, high angular resolution will be helpful in resolving the structure of the source(s). High energy resolution will allow to use fine energy binning in reconstructing the spectrum, as illustrated in Fig. 3, significantly increasing the line detection significance. Search for other lines is also a perspective for GAMMA-400, both in diffuse radiation and from particular objects.

### 2.2 Satellites

A search for dark matter clumps, or satellites, has been a focus of dark matter hunters for a long time. The idea is to look for excessive γ-radiation with extended spatial dimensions, like an extended source, with hard (not power-law) spectrum. The signature to look for is a lack of counterparts in other wavelenths, so unidentified Fermi LAT sources would be the first suspects in the line. The Fermi LAT team checked all their unidentified sources to meet the above crteria and found 2 potential satellite candidates out of 385 unidentified high-latitude sources [11]. However, one of them, 1FGL J1302.3-3255 was found associated in radio observations with a millisecond pulsar, and anoher one, 1FGL J2325.8-4043 was found to have a high probability association with two AGN: 1ES 2322-409 and PKS 2322-411. An upper limit for $<\sigma v>$ is set to $1.95 \times 10^{-24} cm^3 s^{-1}$ for a 100 GeV WIMP annihilation through $b\bar{b}$ channel , still lacking about 2 orders of magnitude to reach predicted WIMP annihilation cross section of $<\sigma v> = 3 \times 10^{-26} cm^3 s^{-1}$.

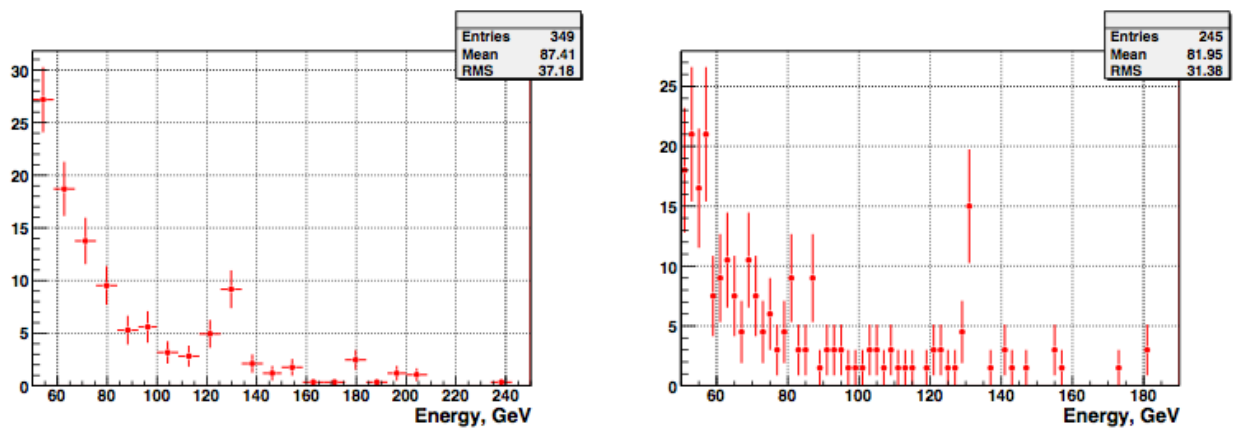

**Figure 3**: Simulation of 130 GeV line detection with the use of toy model. Left panel: simulated for Fermi LAT for 4 years of observation, ∼300 events in total (line + background). Right panel: simulated detection by GAMMA-400 with fewer events, but 10X better energy resolution, reflected in fine energy binning

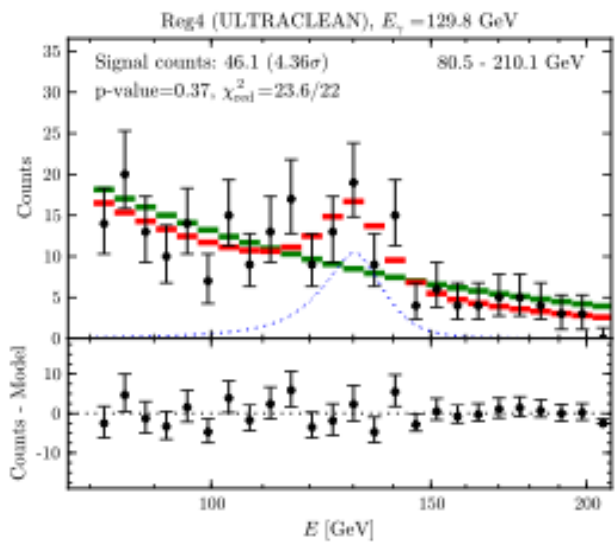

**Figure 2**: Fermi LAT γ-ray line at ∼ 130 GeV [9]

For GAMMA-400 the perspectives will be to investigate with better sensitivity the remaining Fermi LAT and its own unidentified sources. Better energy resolution will allow us to better distinguish between power-law "normal source" and hard DM-originated spectra, potentially increasing the number of satellite candidates. Better angular resolution will allow us to better distinguish between point and extended sources, also potentially increasing the number of satellite candidates. By that time a larger number of available unidentified Fermi LAT sources should also increase the number of satellite candidates.

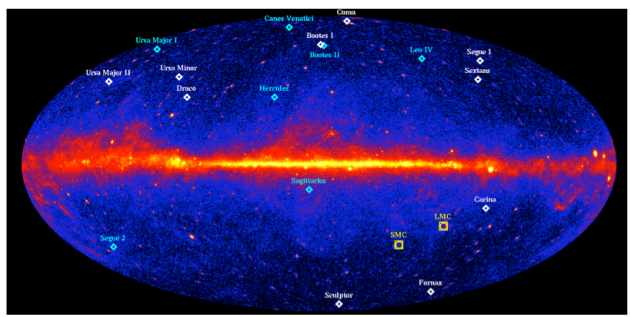

**Figure 4**: Fermi LAT Dwarf Spheroidal Galaxies, used in the stacking DM search [12]

### 2.3 Dwarf Spheroidal Galaxies

The basic idea is to search for excessive γ-ray emission from Dwarf Spheroidal Galaxies (dSph) with large J-factor (line-of-sight integral of the squared DM density). The Fermi LAT team applied a joint (stacking) likelihood analysis to 10 dSph: no dark matter signal was detected (see Fig.4). The upper limit for $<\sigma v>$ was set to $\sim 10^{-26} cm^3 s^{-1}$ at 5 GeV and $5 \times 10^{-23} cm^3 s^{-1}$ at 1 TeV. This is the first result using γ-rays, that rules out the models with the most generic cross section $\sim 3 \times 10^{-26} cm^3 s^{-1}$ for a purely s-wave cross section up to a mass of $\sim$27 GeV for the $b\bar{b}$ channel and up to a mass of $\sim$37 GeV for the $\tau^+ \tau^-$ channel [12]. Improved energy and angular resolution should help GAMMA-400 to reach better sensitivity for this analysis.

### 2.4 Galactic Center

The Galactic Center region is expected to be the strongest source of γ-rays from DM annihilation. Intense background from unresolved sources remains the main problem, assuming that the part of the background created by cosmic ray interactions with matter is much better known and can be accounted for. Potential perspectives for GAMMA-400: having >10 times better angular resolution at high energy, faint sources in the dense GC area can be localized and their radiation can be removed as a background, and a better model of diffuse radiation can be built.

## 3 Summary

- GAMMA-400 will be a very important successor of the Fermi LAT and will provide important observations of γ-rays and cosmic rays in synergy with ground-based γ-ray telescopes and other wavelength instruments. After the end of the Fermi LAT mission, GAMMA-400 will be the only flying γ-ray observatory

- GAMMA-400 main differences from Fermi LAT are $\sim$10 times better angular and energy resolution at energy $> 100$ GeV

- The main objective for GAMMA-400 is to conduct an accurate measurement of the γ-radiation to search for the dark matter smoking gun: γ-ray lines. It will be at least twice as sensitive in this search as Fermi LAT

- Significant contributions to the dark matter constraints will be made with the study of γ-ray satellites (clumps), satellite galaxies (such as dwarf Spheroidal Galaxies), and the Galactic Center. Also important results are expected with cosmic-ray electrons and positrons